\documentclass[twocolumn,english,aps,english,preprintnumbers,groupedaddress,amsmath,amssymb,pra]{revtex4}
\usepackage[T1]{fontenc}
\usepackage[latin9]{inputenc}
\usepackage{amsmath}
\usepackage{amssymb}
\usepackage{esint}

\makeatletter
\@ifundefined{textcolor}{}
{%
 \definecolor{BLACK}{gray}{0}
 \definecolor{WHITE}{gray}{1}
 \definecolor{RED}{rgb}{1,0,0}
 \definecolor{GREEN}{rgb}{0,1,0}
 \definecolor{BLUE}{rgb}{0,0,1}
 \definecolor{CYAN}{cmyk}{1,0,0,0}
 \definecolor{MAGENTA}{cmyk}{0,1,0,0}
 \definecolor{YELLOW}{cmyk}{0,0,1,0}
}

\usepackage{babel}

\usepackage{babel}

\makeatother

\usepackage{babel}
\begin{document}
\title{Comment on \textquotedblleft Gain-assisted superluminal propagation
and rotary drag of photon and surface polaritons\textquotedblright{}}
\author{Bruno Macke}
\author{Bernard S\'{e}gard}
\email{bernard.segard@univ-lille.fr}

\affiliation{Universit\'{e} de Lille, CNRS, UMR 8523, Physique des Lasers, Atomes et
Mol\'{e}cules, F-59000 Lille, France}
\date{May 23, 2019}
\begin{abstract}
In their study of superluminal propagation, rotary drag and surface
polaritons {[}Phys. Rev. A \textbf{96}, 013848 and 049906(E)
(2017){]}, Khan\emph{ et al}. consider a four-level atomic arrangement
with transitions in the optical domain. In fact, the values they give
to the parameters lead to a probe wavelength lying in the decimeter
band and we point out that, in such conditions, all their results
are irrelevant.. 
\end{abstract}
\maketitle
In their study of superluminal propagation, rotary drag and surface
polaritons \cite{art1,Art2}, Khan \emph{et al.} consider a four-level
atomic arrangement with transitions in the optical domain. See Fig.
1(a) in \cite{art1}. On the other hand, they specify in \cite{Art2}
that all the (angular) frequencies are given in units of $\gamma=(2\pi)\times1\,\mathrm{MHz}$
and that the probe frequency $\nu_{p}=1000\gamma$. The corresponding
wavelength is thus $\lambda_{p}=30\,\mathrm{cm}$ (in the decimeter
band!). As shown in the following this invalidates all the results
given in \cite{art1,Art2}.

As correctly given in \cite{art1}, the electric susceptibility for
the probe reads, in SI units,:
\begin{equation}
\chi=\frac{2N\left|\wp_{ac}\right|^{2}\rho_{ac}}{\varepsilon_{0}\hbar\Omega_{p}}\label{eq:un}
\end{equation}
where $N$ is the atomic number density, $a$ ($c$) is the upper
(lower) level of the probe transition, $\wp_{ac}$ ($\rho_{ac}$)
is the corresponding matrix element of the dipole moment (of the density
operator) and $\Omega_{p}$ is the Rabi (angular) frequency of the
probe. Expressing the susceptibility as a function of the probe wavelength
as made to obtain Eq. (5) in \cite{art1,Art2} can be achieved by
introducing the Einstein\textquoteright s coefficient $A_{ac}$ associated
with the transition $a\rightarrow c$. From its expression given in
\cite{Art3}, we get:
\begin{equation}
\left|\wp_{ac}\right|^{2}=\left(\frac{3\lambda_{p}^{3}}{8\pi^{2}}\right)\hbar\varepsilon_{0}A_{ac}\label{eq:deux}
\end{equation}
and finally
\begin{equation}
\chi=\left(\frac{3N\lambda_{p}^{3}}{32\pi^{3}}\right)\left(\frac{8\pi A_{ac}}{\Omega_{p}}\right)\rho_{ac}.\label{eq:trois}
\end{equation}
The expression $\chi=\left(\frac{3N\lambda_{p}^{3}}{32\pi^{3}\Omega_{p}}\right)\rho_{ac}$
given by Eq. (5) in \cite{Art2} thus holds only if $\Omega_{p}$
is expressed in units of $8\pi A_{ac}$. According to the above choice
of $\gamma$ as unit of (angular) frequency, this implies that $8\pi A_{ac}=\gamma$.

It is specified in \cite{Art2} that \textquotedblleft the susceptibility
and group index plotted versus probe detuning have units of $2N\left|\wp_{ac}\right|^{2}/(\varepsilon_{0}\hbar)$\textquotedblright .
As shown in Eq.(\ref{eq:un}), this quantity has the dimension of
an angular frequency and, for consistency, it should also be expressed
in units of $\gamma$. It then reads $u_{\chi}=2N\left|\wp_{ac}\right|^{2}/(\varepsilon_{0}\hbar\gamma)$
and, taking into account the above relations,
\begin{equation}
u_{\chi}=\frac{3N\lambda_{p}^{3}}{32\pi^{3}}\label{eq:quatre}
\end{equation}

For wavelengths $\lambda_{p}$ in the visible domain and typical values
of the atomic number density $N$, the susceptibility unit $u_{\chi}$
given by Eq. (\ref{eq:quatre}) is in the order of $3\times10^{-3}$.
On the other hand, for $\lambda_{p}=30\,\mathrm{cm}$ with $N=5\times10^{12}\,\mathrm{cm^{-3}}$
as considered in \cite{Art2}, this unit rises to $u_{\chi}\approx4\times10^{14}$.
Figure 2 in \cite{Art2} shows that the peak value of the \emph{relative}
susceptibility $\chi/u_{\chi}$ can exceed $5\times10^{-3}$. The
corresponding \emph{absolute} susceptibility $\chi$ is then in the
order of $10^{12}$. \emph{Such values are meaningless}.

Although this point is less important, we note that, in SI units,
the refractive index reads $n=\sqrt{1+\chi}$ and not $n=\sqrt{1+4\pi\chi}$
as used in \cite{art1} to determine the group index. Anyway the approximation
$n\approx1+2\pi\chi$ also made to obtain Eq. (6) in \cite{art1}
fails when $\left|\chi\right|\gg1$.

Without examining in detail the parts of \cite{art1,Art2} devoted
to rotary drag and surface polaritons, we remark that these phenomena
occur when the sample thickness $L$ is large compared to the probe
wavelength $\lambda_{p}$ . According to \cite{Art2}, $L=10\,\mathrm{cm}$
and this condition is far from being fulfilled since this thickness
is only one third of the probe wavelength. By the way, we also note
the incompatibility of the figures 3(b) and 4 in \cite{Art2} which show
rotary drags, respectively, in the order of $10^{-2}$ and $10^{-7}$
rad.

Khan \emph{et al.} support their choice of the ratio $\nu_{p}/\gamma=1000$
by referring to a paper on the phase control of light velocity \cite{Art4}.
The same ratio was actually considered in this paper but without specifying
the absolute value of the frequencies. We, however, point out that,
for a probe frequency in the visible domain, this ratio leads to lifetimes
of the excited atomic levels which are fully unrealistic (in the subpicosecond
domain).

Independently of the above criticisms, we remark that, quite generally,
large negative group delays are not a sufficient condition to observe
visible effects of superluminal propagation. A convincing demonstration
of such effects would have required a comparison of the transmitted
and incident pulses, which is not made in \cite{art1,Art2}.

This work has been partially supported by the Minist\`{e}re de l'Enseignement
Sup\'{e}rieur, de la Recherche et de l'Innovation, the Conseil R\'{e}gional
des Hauts de France and the European Regional Development Fund (ERDF)
through the Contrat de Projets \'{E}tat-R\'{e}gion (CPER) 2015--2020, as
well as by the Agence Nationale de la Recherche through the LABEX
CEMPI project (Project No. ANR-11-LABX-0007).

\end{document}